# Giant anomalous Hall effect in a ferromagnetic Kagomé-lattice semimetal

Enke Liu[1,2][*], Yan Sun[1][*], Nitesh Kumar[1], Lukas Müchler[3], Aili Sun[1], Lin Jiao[1], Shuo-Ying Yang[4], Defa Liu[4], Aiji Liang[5,6], Qiunan Xu[1], Johannes Kroder[1], Vicky Süß[1], Horst Borrmann[1], Chandra Shekhar[1], Zhaosheng Wang[7], Chuanying Xi[7], Wenhong Wang[2], Walter Schnelle[1], Steffen Wirth[1], Yulin Chen[5,8], Sebastian T. B. Goennenwein[9] and Claudia Felser[1][*]

[1] Max-Planck Institute for Chemical Physics of Solids, Dresden, Germany

[2] Institute of Physics, Chinese Academy of Sciences, Beijing, China

[3] Department of Chemistry, Princeton University, Princeton, New Jersey, USA

[4] Max Planck Institute of Microstructure Physics, Halle, Germany

[5] School of Physical Science and Technology, ShanghaiTech University, Shanghai, China

[6] Advanced Light Source, Lawrence Berkeley National Laboratory, Berkeley, California, USA

[7] High Magnetic Field Laboratory, Chinese Academy of Sciences, Hefei, Anhui, China

[8] Clarendon Laboratory, Department of Physics, University of Oxford, Parks Road, Oxford, UK

[9] Institut für Festkörper- und Material Physik, Technische Universität Dresden, Dresden, Germany

[*] e-mail: ekliu@iphy.ac.cn; Yan.Sun@cpfs.mpg.de; Claudia.Felser@cpfs.mpg.de

**ABSTRACT**

Magnetic Weyl semimetals with broken time-reversal symmetry are expected to generate strong intrinsic anomalous Hall effects, due to their large Berry curvature. Here, we report a magnetic Weyl semimetal candidate $Co_3Sn_2S_2$ with a quasi-two-dimensional crystal structure consisting of stacked Kagomé lattices. This lattice provides an excellent platform for hosting exotic quantum topological states. We observe a negative magnetoresistance that is consistent with the chiral anomaly expected from the presence of Weyl fermions close to the Fermi level. The anomalous Hall conductivity is robust against both increased temperature and charge conductivity, which corroborates the intrinsic Berry-curvature mechanism in momentum space. Owing to the low carrier density in this material and the significantly enhanced Berry curvature from its band structure, the anomalous Hall conductivity and the anomalous Hall angle simultaneously reach 1130 $\Omega^{-1}$ cm$^{-1}$ and 20%, respectively, an order of magnitude larger than typical magnetic systems. Combining the Kagomé-lattice structure and the long-range out-of-plane ferromagnetic order of $Co_3Sn_2S_2$, we expect that this material is an excellent candidate for observation of the quantum anomalous Hall state in the two-dimensional limit.

*Electrical transport measurements reveal that $Co_3Sn_2S_2$ is probably a magnetic Weyl semimetal, and hosts the highest simultaneous anomalous Hall conductivity and anomalous Hall angle. This is driven by the strong Berry curvature near the Weyl points.*

The anomalous Hall effect (AHE) is an important electronic transport phenomenon[1]. It can arise because of two qualitatively different microscopic mechanisms: Extrinsic processes due to scattering effects, and an intrinsic mechanism connected to the Berry curvature[1-5]. The large Berry curvature comes from the entangled Bloch electronic bands with spin-orbit coupling when the spatial-inversion or time-reversal symmetry of the material is broken[6,7]. The quantum AHE in two-dimensional (2D) systems is determined solely by this intrinsic contribution[8,9]. It manifests itself as a quantized anomalous Hall conductance due to the presence of a bulk gap in combination with dissipationless edge states[10-13]. A magnetic Weyl semimetal with broken time-reversal symmetry can be interpreted as a stacked heterostructure of such quantum anomalous Hall insulator layers[14,15], where the coupling between the layers closes the bulk band gap at isolated Weyl nodes. At these Weyl nodes, the Berry curvature is enhanced while the carrier density vanishes[2-4,16,17]. This suggests that an intrinsic large anomalous Hall conductivity and a large anomalous Hall angle can be expected in such systems.

To date, a number of promising candidates for magnetic Weyl semimetals have been proposed, including $Y_2Ir_2O_7$ ([18]), $HgCr_2Se_4$ ([19]), and certain $Co_2$-based Heusler compounds[20-22]. The experimental identifications for this Weyl phase in these systems are also on the way. Indeed, an anomalous Hall angle of ~16% was recently observed at low temperatures in the magnetic-field induced Weyl semimetal GdPtBi ([23]). However, a finite external magnetic field is mandatory to make GdPtBi a Weyl semimetal. Therefore, the search for intrinsic magnetic Weyl semimetals with Weyl nodes close to the Fermi level is not only an efficient strategy to obtain materials exhibiting both a high anomalous Hall conductivity and large anomalous Hall angle, but also important for a comprehensive understanding of the Weyl topological effects on the AHE in real materials.

The Kagomé lattice has become one of the most fundamental models for exotic topological states in condensed matter physics. In particular, the Kagomé lattice with out-of-plane magnetization is an excellent platform for quantum anomalous Hall effect[24,25]. Thus, it provides an effective guiding principle for realizing magnetic Weyl semimetals via stacking[14,15]. Although a Dirac dispersion with a finite spin-orbit-coupling induced gap has recently been observed in a Kagomé-lattice metal[26], the Weyl phase in a magnetic Kagomé material still remains elusive. Here, we report a time-reversal-symmetry-breaking Weyl semimetal in the magnetic Kagomé-lattice compound $Co_3Sn_2S_2$ with out-of-plane ferromagnetic order, and demonstrate both a large intrinsic anomalous Hall conductivity (1130 $\Omega^{-1}$ cm$^{-1}$) and giant anomalous Hall angle (20%).

$Co_3Sn_2S_2$, a Shandite compound, is known to be a ferromagnet with a Curie temperature ($T_C$) of 177 K and a magnetic moment of 0.29 $\mu_B$/Co [27-29]. Magnetization measurements have shown that the easy axis of the magnetization lies along the *c*-axis[30], while photoemission measurements and band structure calculations revealed that below $T_C$, $Co_3Sn_2S_2$ exhibits Type-IA half-metallic ferromagnetism in which spin-minority states are gapped[31,32]. Figure 1 summarizes the structural and electronic properties of $Co_3Sn_2S_2$. As shown in Fig. 1a, $Co_3Sn_2S_2$ crystallizes in a rhombohedral structure of the space group, *R*-3*m* (no. 166) [27]. The crystal possesses a long-range quasi-2D $Co_3Sn$ layer sandwiched between sulphur atoms, with the magnetic cobalt atoms arranged on a Kagomé lattice in the *a-b* plane in the hexagonal representation of the space group. Owing to the strong magnetic anisotropy, this material shows a quasi-2D nature of magnetism[30]. Our magnetization measurements revealed a quite low saturation field (~ 0.05 T) along the *c*-axis and an extremely high one (> 9 T) in the *a-b* plane, confirming a dominantly out-of-plane magnetization in $Co_3Sn_2S_2$ (see Supplementary Information). By itself, the dimensional restriction of the out-of-plane magnetization may be responsible for some of the interesting electronic and magnetic properties of this compound. We discuss band structure calculations of $Co_3Sn_2S_2$ with spin polarization along the *c*-axis. The calculated magnetic moments without and with spin-orbit coupling are 0.33 and 0.30 $\mu_B$/Co, respectively, which are very close to the experimental values of 0.29 $\mu_B$/Co obtained from neutron diffraction[29], 0.31 $\mu_B$/Co from magnetization measurement[30], and 0.30 $\mu_B$/Co from our measurement (see Supplementary Information). As expected, the calculation including spin-orbit coupling yields a more accurate result.

The band structures of $Co_3Sn_2S_2$ calculated without and with spin-orbit coupling are shown in Fig. 1b. The bands corresponding to the spin-down channel are insulating in character with a gap of 0.35 eV, while the spin-up channel crosses the Fermi level and thus has metallic character. This half-metallic behaviour is consistent with the results of previous studies on this compound[30-32]. Furthermore, for the spin-up states, we observe linear band crossings along the Γ–L and L–U paths, just slightly above and below the Fermi energy, respectively. For finite spin-orbit coupling, these linear crossings open small gaps with band anti-crossings, and make this compound semimetal-like. The relatively small Fermi surfaces (Fig. 1c), showing the coexistence of holes and electrons, further corroborate the semi-metallic character in this compound. This calculated band structure is in good agreement with our angle-resolved photoemission spectroscopy (ARPES) measurements (see Supplementary Information). When these results are considered in connection with the ferromagnetism of $Co_3Sn_2S_2$ (27-30), they suggest that a time-reversal-symmetry-breaking Weyl semimetal phase might be hidden in this compound.

In order to confirm this prediction, single crystals of $Co_3Sn_2S_2$ were grown for further experimental investigations (see Methods and Supplementary Information). The high quality of the crystals was confirmed by structure refinements based on single-crystal X-ray diffraction and topographic images of the hexagonal lattice array obtained using scanning tunneling microscopy (see Supplementary Information). As shown in Fig. 1d, the longitudinal electric resistivity ($\rho$) decreases with decreasing temperature, showing a kink at $T_C$ = 175 K and a moderate residual resistivity of ~50 μΩ cm at 2 K. In a high field of 9 T, a negative magnetoresistance appears around the Curie temperature owing to the spin-dependent scattering in magnetic systems. At low temperatures, the MR increases and becomes positive (Fig. 1d). This behaviour is further demonstrated by the field dependent resistance (Fig. 1e). Importantly, the positive magnetoresistance shows no signature of saturation even up to 14 T, which is typical of a semi-metal[33,34]. The notable non-linear field dependence of the Hall resistivity ($\rho_H$) (Fig. 1f) further indicates the coexistence of hole and electron carriers at 2 K, which is in good agreement with our band structure calculations (Figs. 1b and 1c). By using the semiclassical two-band model[35], we extract the carrier densities of holes ($n_h$ ~ $9.3 \times 10^{19}$ cm$^{-3}$) and electrons ($n_e$ ~ $7.5 \times 10^{19}$ cm$^{-3}$) of our $Co_3Sn_2S_2$ samples. These relatively low carrier densities and a near compensation of charge carriers further confirm the semi-metallicity of $Co_3Sn_2S_2$.

In order to further analyze the topological character of $Co_3Sn_2S_2$ suggested by Fig. 1b, we now consider the linear band crossings in more detail. The space group *R*-3*m* of $Co_3Sn_2S_2$ has one mirror plane $M_{010}$. Without spin-orbit coupling, the interaction between spin-up and spin-down states is ignored and the mirror plane is a high symmetry plane of the Hamiltonian. Thus, as they are protected by this mirror symmetry, the linear band crossing identified in Fig. 1b form a nodal ring in the mirror plane based on the band inversion, as shown in Fig. 2a. Moreover, the linear crossings between the L–Γ and L–U paths are just single points on the ring. When the $C_{3z}$-rotation and inversion symmetries of the material are considered, one finds a total of six nodal rings in the Brillouin zone, as shown schematically in Fig. 2b.

Upon taking spin-orbit coupling into account, the spin $s_z$ is no longer a good quantum number and the mirror symmetry of the Hamiltonian is broken, which causes the linear crossings of the nodal lines to split, as presented in Fig. 2c. Interestingly, one pair of linear crossing points remains in the form of Weyl nodes along the former nodal line. These two Weyl nodes act as a monopole sink and source of Berry curvature (see Supplementary Information) and possess opposite topological charges of +1 and −1, respectively. In total, there are three such pairs of Weyl nodes in the first Brillouin zone due to the inversion and $C_{3z}$-rotation symmetries of the crystal, and their distribution is presented in Fig. 2b. It is important to emphasize that the Weyl nodes in $Co_3Sn_2S_2$ are only 60 meV above the charge neutrality point, which is much closer to the Fermi energy than previously proposed magnetic Weyl semimetals. These Weyl nodes and non-trivial Weyl nodal rings together make this material exhibit a simple topological band structure around the $E_F$. It is thus easy to further observe the surface Fermi arcs[36]. As a result, the Weyl node-dominated physics in $Co_3Sn_2S_2$ should be prominent and easy to detect in experiments.

We now address the AHE response of $Co_3Sn_2S_2$ that can be expected from the particular band structure properties outlined above. In order to obtain a complete topological character, we integrated the Berry curvature $\Omega_{yx}^z(k)$ along $k_z$ in the Brillouin zone. Our results reveal two main types of hot spots for the integrated Berry curvature: One that is located around the Weyl nodes, and the other near the edge of the nodal lines (see Fig. 2d). To investigate the origin of the hot spot of the Berry curvature distribution, we choose the $k_y = 0$ plane, which includes two nodal rings and two pairs of Weyl nodes, as shown in Fig. 2e. We note that hot spots of the integrated Berry curvature are primarily determined by the shape of the nodal lines, and both types of hot spots observed here originate from the nodal-line-like band anti-crossing behaviour. Along the nodal ring, the component of the Berry curvature parallel to $k_z$ leads to the larger hot spot we observe, while a different part around the Weyl node contributes to the smaller hot spot. Owing to the band anti-crossing behaviour and the position of the six Weyl nodal rings around the Fermi level, the calculated Berry curvature is clean and large, which should yield fascinating spin-electronic transport behaviours including a large intrinsic AHE[3].

The energy dependent anomalous Hall conductivity ($\sigma_{yx}$) calculated from the Berry curvature is shown in Fig. 2f. As one can see from the figure, a large peak in $\sigma_{yx}$ appears around $E_F$ with a maximum of 1100 $\Omega^{-1}$ cm$^{-1}$. Since the $\sigma_{yx}$ depends on the location of the Fermi level (see Eq. (3), Methods), it usually changes sharply as a function of energy. However, the peak in $\sigma_{yx}$ in Fig. 2f stays above 1000 $\Omega^{-1}$ cm$^{-1}$ within an energy window of 100 meV below $E_F$. Therefore, we expect to observe a high $\sigma_{yx}$ in experiments for charge neutral or slightly *p*-doped $Co_3Sn_2S_2$ samples. We also consider the non-collinear magnetic structure of the Kagomé lattice in $Co_3Sn_2S_2$. During spin tilting away from *c*-axis, the calculated $\sigma_{yx}$ always stays above 1000 $\Omega^{-1}$ cm$^{-1}$. The existence of Weyl nodes and large anomalous Hall conductivity are robust against the change of the magnetic structure of $Co_3Sn_2S_2$ (see Supplementary Information).

A Weyl semimetal is expected to exhibit the so-called chiral anomaly[37] in transport, when the conservation of chiral charges is violated in case of a parallel magnetic and electric field, as shown in Fig. 3a. We measured the impact of magnetic field orientation on longitudinal resistivity at 2 K (Fig. 3b). For $B \perp I$ ($\theta = 90°$), the positive unsaturated magnetoresistance (also see Fig. 1e) is observed. The magnetoresistance decreases rapidly with decreasing $\theta$. A clear negative magnetoresistance appears when $B // I$ ($\theta = 0°$) which again does not saturate up to 14 T. As an equivalent description of the magnetoresistance, the magnetoconductance is shown in Fig. 3c. In the parallel case ($B // I$), the positive magnetoconductance can be well described by a near parabolic field dependence[38], ~ $B^{1.9}$, up to 14 T (Inset of Fig. 3c). In this case, the charge carriers are pumped from one Weyl point to the other one with opposite chirality, which leads to an additional contribution to the conductance, resulting in a negative magnetoresistance[37,38]. The chiral anomaly evident from Fig. 3 represents an important signature of the Weyl fermions in $Co_3Sn_2S_2$.

Our transport measurements further verify the strong AHE induced by the Weyl band topology. An out−of−plane configuration of $I$ // $x$ //[$2\bar{1}\bar{1}0$] and $B$ // $z$ // [0001] was applied in these measurements (see Fig. 1d and Methods). As we observe in Fig. 4a, the anomalous Hall conductivity ($\sigma_H^A$) (see Methods) shows a high value of 1130 $\Omega^{-1}$ cm$^{-1}$ at 2 K, which is in very good agreement with our predicted theoretical value ($\sigma_{yx}$, Fig. 2f). We also studied the in−plane case ($I$ // $x$ // [$2\bar{1}\bar{1}0$] and $B$ // $y$ // [$01\bar{1}0$]), for which the AHE disappears (not shown), due to strong magnetic and Berry-curvature anisotropies. Moreover, at temperatures below 100 K, for the out−of−plane case $\sigma_H^A \sim 1000$ $\Omega^{-1}$ cm$^{-1}$ is revealed to be independent of temperature (see also the inset of Fig. 4a, and note the logarithmic vertical axis). This robust behaviour against temperature indicates that the AHE is not governed by scattering events in the system. In addition, $\sigma_H$ shows rectangular hysteresis loops with very sharp switching (Fig. 4b), and the coercive field is seen to increase with decreasing temperature, resulting in a value of 0.33 T at 2 K (also see Supplementary Information). As is evident from the figure, a large remanent Hall effect at zero field is observed in this material.

We plot $\rho_H^A$ as a function of temperature in Fig. 4c. A large peak in $\rho_H^A$ with a maximum of 44 $\mu\Omega$ cm appears at 150 K. When $\sigma_H^A$ is plotted against $\sigma$, as presented in Fig. 4d, we also find that $\sigma_H^A$ is nearly independent of $\sigma$ (i.e., $\sigma_H^A \sim (\sigma)^0$ = constant) for temperatures below 100 K, as is expected for an intrinsic AHE in the framework of the unified model for AHE physics[39,40] (see Supplementary Information for more details). This independence of $\rho_H^A$ with respect to both $T$ and $\rho$ indicates that the AHE only originates from the intrinsic scattering-independent mechanism, and is thus dominated by the Berry curvature in momentum space[1]. This scaling behaviour is well consistent with our first-principles calculations and also provides another important signature for the magnetic Weyl fermions in $Co_3Sn_2S_2$.

In addition to a large $\sigma_H^A$, and arguably more importantly, the magnetic Weyl semimetal $Co_3Sn_2S_2$ also features a giant anomalous Hall angle that can be characterized by the ratio of $\sigma_H^A/\sigma$. The temperature dependence of the $\sigma_H^A/\sigma$ is shown in Fig. 5a. With increasing temperature, the $\sigma_H^A/\sigma$ first increases from 5.6% at 2 K, reaching a maximum of ~ 20% around 120 K, before decreasing again as the temperature increases above $T_C$. The contour plot of $\sigma_H/\sigma$ with respect to *B* and *T* is depicted in Fig. 5b, and makes it intuitively clear that a giant Hall angle appears between 75 – 175 K irrespective of the magnetic field magnitude. This can be straightforwardly understood by considering that the $\sigma_H^A$ arises from the Berry curvature of the occupied states. The band topology of these states is basically unaffected by the small energy scale of thermal excitations up to room temperature[41]. In other words, the topologically protected $\sigma_H^A$ is relatively robust against temperature. In contrast, the Weyl-node-related charge conductivity ($\sigma$) is sensitive to temperature due to electron–phonon scattering[42]. These behaviours are also shown in Fig. 5a. Therefore, the $\sigma_H^A/\sigma$ is expected to increase with increasing temperature in a wide temperature range below $T_C$. The semi-metallicity (low carrier density and low charge conductivity) largely improves the value of $\sigma_H^A/\sigma$ in this system.

When compared to previously reported results for other AHE materials (see Fig. 5c), the value of anomalous Hall angle in $Co_3Sn_2S_2$ observed in this work is seen to be the largest by quite a prominent margin. For most of these materials — formed mainly of ferromagnetic transition metals and alloys — the anomalous Hall conductivities originate from topologically trivial electronic bands. A typical feature of these materials is that both the $\sigma_H^A$ and the $\sigma$ are either large or small and therefore the $\sigma_H^A/\sigma$ of these materials typically cannot be large. While the magnetic-field-induced Weyl semimetal GdPtBi has a large $\sigma_H^A/\sigma$ of 16%, its $\sigma_H^A$ is very small, and moreover, it requires an external field to induce the Weyl phase[23]. In contrast, owing to the non-trivial Berry curvature and the Weyl semi-metallic character, the Kagomé-lattice $Co_3Sn_2S_2$ possesses both a large $\sigma_H^A$ and giant $\sigma_H^A/\sigma$ simultaneously and at zero magnetic field, which promotes this system to quite a different position among the known AHE materials. As a consequence, a large anomalous Hall current can be expected in thin films of this material that may even reach the limit of a quantized AHE with dissipationless quantum Hall edge states[13,24,43,44]. In more general terms, a clean topological band structure induces both a large anomalous Hall conductivity and giant anomalous Hall angle (as demonstrated here for the Weyl semimetal $Co_3Sn_2S_2$), and so can be seen as a guide for the realization of strong AHE in (half-metallic) magnetic topological Weyl semimetals.

In summary, $Co_3Sn_2S_2$ is a Weyl semimetal candidate derived from a ferromagnetic Kagomé lattice. It is the first material that hosts both a large anomalous Hall conductivity and a giant anomalous Hall angle that originate from Berry curvature. This compound is an ideal candidate for developing a quantum anomalous Hall state due to its long-range quasi-2D out-of-plane ferromagnetic order and simple electronic structure near the Fermi energy. Moreover, it is straightforward to grow large, high-quality, single crystals, which makes $Co_3Sn_2S_2$ and the Shandite family an excellent platform for comprehensive studies on topological electron behaviour. Our work motivates the study of the strong anomalous Hall effect based on magnetic Weyl semimetals, and establishes the ferromagnetic Kagomé lattice Weyl semimetals as a key class of materials for fundamental research and applications connecting topological physics[45-48] and spintronics[49,50].

**Acknowledgements**

This work was financially supported by the European Research Council (ERC) Advanced Grant (No. 291472) 'IDEA Heusler!' and ERC Advanced Grant (No. 742068) 'TOPMAT'. E.L. acknowledges supports from Alexander von Humboldt Foundation of Germany for his Fellowship and from National Natural Science Foundation of China for his Excellent Young Scholarship (No. 51722106).

**Author contributions**

The project was conceived by E.L. and C.F. E.L. grew the single crystals and performed the structural, magnetic and transport measurements with the assistances from A.S., J.K., S.Y., V.S., H.B., N.K. and W.S. The STM characterizations were performed by L.J. and S.W. The ARPES measurements were conducted by D.L., A.L. and Y.C. The static high magnetic field measurements were performed and analyzed by Z.W., C.X., N.K., C.S. and L.J. The theoretical calculations were carried out by Y.S., L.M. Q.X. and E.L. All the authors discussed the results. The paper was written by E.L., Y.S. and S.T.B.G. with feedback from all the authors. The project was supervised by C.F.

**Competing financial interests**

The authors declare no competing financial interests.

**Methods**

**Single-crystal growth**. The single crystals of $Co_3Sn_2S_2$ were grown by self-flux methods with Sn as flux or with the congruent composition in a graphite crucible sealed in a quartz tube (see Supplementary Information). The stoichiometric samples (Co : Sn : S = 3 : 2 : 2) were heated to 1000 °C over 48 hours and kept there for 24 hours before being slowly cooled to 600 °C over 7 days. The samples were kept at 600 °C for 24 hours to obtain the homogeneous and ordered crystals. The compositions of crystals were checked by energy dispersive X-ray spectroscopy. The crystals were characterized by powder X-ray diffraction as single phase with a Shandite-type structure. The lattice parameters at room temperature are $a$ = 5.3689 Å and $c$ = 13.176 Å. The single crystals and orientations were confirmed by a single-crystal X-ray diffraction technique.

**Scanning tunneling microscopy (STM).** Topographic images of the crystal surface were characterized by a cryogenic STM, taken at conditions of $T$ = 2.5 K, a bias voltage of $V_b$ = 100 mV, and a tunnel current of $I_t$ = 500 pA. The sample was cleaved *in situ* ($p \leq 2 \times 10^{-9}$ Pa) at 20 K. The high quality of the single crystals was confirmed by STM (see Supplementary Information).

**Magnetization measurements.** Magnetization measurements were carried out on oriented crystals with the magnetic field applied along both the *a* and *c* axes using a vibrating sample magnetometer (MPMS 3, Quantum Design). The results show an extremely strong magnetic anisotropy in $Co_3Sn_2S_2$ (see Supplementary Information).

**Out-of-plane transport measurements.** The out-of-plane transport measurements on longitudinal charge and Hall resistivities, with $B // z //[0001]$ and $I // x //[2\bar{1}\bar{1}0]$, were performed on a PPMS 9 (Quantum Design) using the low-frequency alternating current (ACT) option. The standard four-probe method was used to measure the longitudinal electrical resistivity, while for the Hall resistivity measurements, the five-probe method was used with a balance protection meter to eliminate possible magnetoresistance signals. The charge and Hall resistivities were measured alternatively at each temperature.

**Angle dependent longitudinal electric resistivity.** The angle dependence of longitudinal electric resistivity was measured on PPMS DynaCool (Quantum Design) using the DC Resistivity Option. For the angle-dependent measurements, $B // \theta$ and $I // x //[2\bar{1}\bar{1}0]$, while $\theta$ is the angle with respect to $x //[2\bar{1}\bar{1}0]$. The currents were always applied along the *a*-axis, e.g. $I // x // [2\bar{1}\bar{1}0]$ (*a* axis = *x* axis). Different crystals, grown by two self-flux methods and with different RRR ($\rho_{300K}/\rho_{2K}$) values, were used in this study.

**Analysis of Hall effect and semi-metallicity.** At high temperatures (50 K < $T$ < $T_C$), the Hall signal shows a linear field-dependent behaviour after saturation. At low temperatures ($T$ < 50 K), a notable non-linear field dependence of the Hall resistivity is observed, indicating the existence of two types of carriers (electrons and holes). The electron carriers appear at low temperatures. The single-band and two-band models were thus applied to extract the pure anomalous Hall resistivity, carrier densities and mobilities, for high-temperature and low-temperature cases, respectively.

The anomalous Hall conductivity was calculated by

$$\sigma_H^A = -\rho_H^A/((\rho_H^A)^2 + \rho^2) \tag{1}$$

Here $\rho_H^A$ is the anomalous Hall resistivity at zero field; $\rho$ is the longitudinal resistivity at zero field.

The two-band model[35] was applied to extract the densities of both carriers at low temperatures.

$$\sigma(B) = \frac{n_h e \mu_h}{1+\mu_h^2 B^2} + \frac{n_e e \mu_e}{1+\mu_e^2 B^2}$$

$$\sigma_H(B) = \frac{n_h e \mu_h^2 B}{1+\mu_h^2 B^2} + \frac{n_e e \mu_e^2 B}{1+\mu_e^2 B^2} \tag{2}$$

Here $B$ is applied magnetic field, $\sigma(B)$ is longitudinal charge conductivity, $\sigma_H(B)$ is anomalous Hall effect, $n_h$ is carrier concentration of holes and $\mu_h$ is carrier mobility of holes, $n_e$ is carrier concentration of electrons and $\mu_e$ is carrier mobility of electrons.

**Longitudinal magnetoresistance in static high magnetic fields.** The field-dependent longitudinal magnetoresistance was measured in static high magnetic fields up to 37 T, by a standard four-probe method in a 3He cryostat with $B$ // $c$-axis, using a hybrid magnet at the High Magnetic Field Laboratory, Chinese Academy of Sciences. The current was 5 mA at a frequency of 13.7 Hz applied by a Keithley 6221. The voltage was measured by a SR830 Lock-In Amplifier. The Shubnikov-de Haas (SdH) quantum oscillations of magnetoresistivity was observed above 17 T in the present crystal. A cubic polynomial background was subtracted from the resistivity data. For the fast Fourier transform, Hanning window was applied in the Origin software.

**Density functional theory (DFT) calculations.** The electronic structure calculations were performed based on the DFT using the Vienna *ab-initio* simulation package (VASP)[51]. The exchange and correlation energies were considered in the generalized gradient approximation (GGA), following the Perdew–Burke–Ernzerhof parametrization scheme[52]. We have projected the Bloch wave functions into Wannier functions[53], and constructed the tight binding model Hamiltonian based on the Wannier functions. The anomalous Hall conductivity and Berry curvature was calculated by the Kubo formula approach in the linear response and clean limit[4]:

$$\sigma_{yx}^z(E_F) = e^2\hbar(\frac{1}{2\pi})^3 \int_{\vec{k}} d\vec{k} \sum_{E(n,\vec{k})<E_F} f(n,\vec{k})\Omega_{n,yx}^z(\vec{k})$$

$$\Omega_{n,yx}^z(\vec{k}) = \text{Im}\sum_{n'\neq n} \frac{<u(n,\vec{k})|\hat{v}_y|u(n',\vec{k})><u(n',\vec{k})|\hat{v}_x|u(n,\vec{k})> - (x \leftrightarrow y)}{(E(n,\vec{k})-E(n',\vec{k}))^2} \tag{3}$$

where $f(n,\vec{k})$ is the Fermi-Dirac distribution, $E(n,\vec{k})$ is the eigen-value of the *n-th* eigen-state of $|u(n,\vec{k})>$ at $\vec{k}$ point, and $\hat{v}_{x(y)} = \frac{1}{\hbar}\frac{\partial H(\vec{k})}{\partial k_{x(y)}}$ is the velocity operator. The numerical integration was performed using a $501 \times 501 \times 501$ *k*-grid. The Fermi surfaces were calculated by a *k*-grid of $120 \times 120 \times 120$ from the tight binding model Hamiltonian, and the frequencies of electron oscillations were calculated from the extremal cross-section areas of the Fermi surface perpendicular to the applied magnetic field (see Supplementary Information).

**Angle-resolved photoemission spectroscopy (ARPES).** ARPES measurements on single crystals were performed at beamline BL5-2 of the Stanford Synchrotron Radiation Lightsource, SLAC National Accelerator Laboratory, and the beamline BL10.0.1 of the Advanced Light Source (ALS). The data were recorded by a Scienta R4000 analyzer at *p* = $4 \times 10^{-9}$ Pa at 20 K in both facilities. The total convolved energy and angle resolutions were *E* = 10 to 20 meV and *θ* = 0.2 °, respectively. Good agreements of the Fermi surfaces and energy dispersions from ARPES measurements and DFT calculations are also obtained (see Supplementary Information).

**Data availability**. The data that support the plots within this paper and other findings of this study are available from the corresponding author upon reasonable request.

**CAPTIONS OF FIGURES**

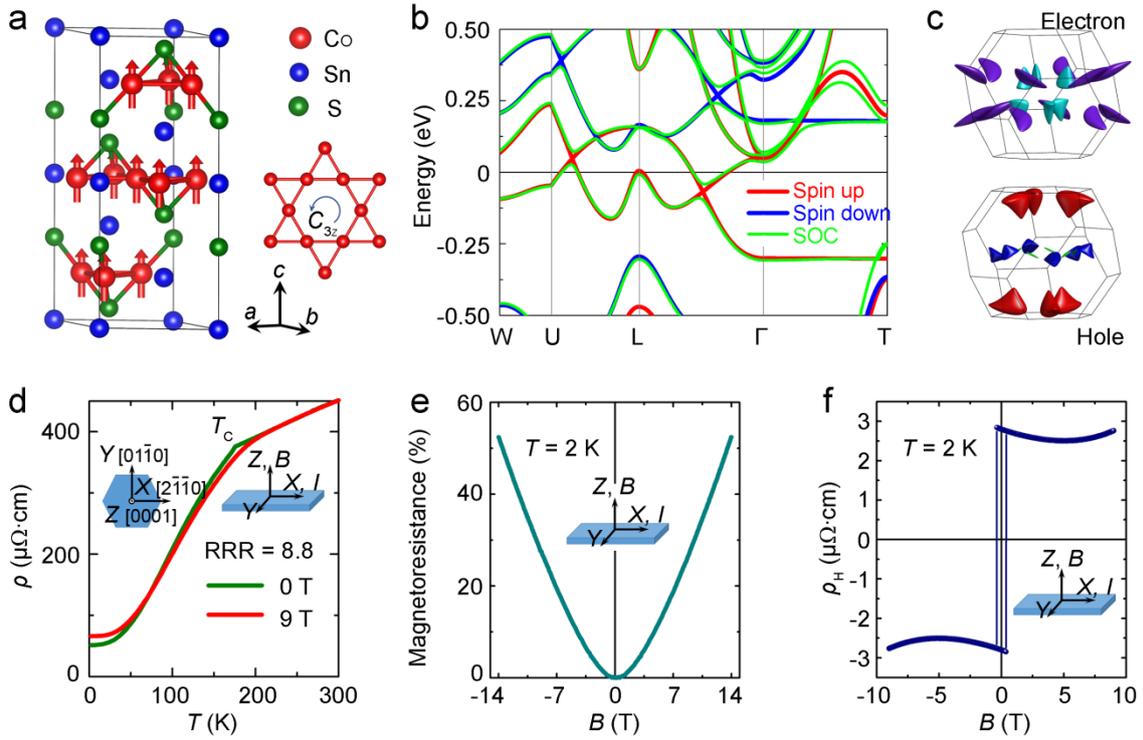

**Figure 1 | Crystal and electronic structures of $Co_3Sn_2S_2$ and the measured electric resistivity**. **a,** Unit cell in a hexagonal setting. The cobalt atoms form a ferromagnetic Kagomé lattice with a $C_{3z}$-rotation. The magnetic moments are shown along the *c*-axis. **b,** Energy dispersion of electronic bands along high-symmetry paths without and with spin-orbit coupling, respectively. "SOC" denotes "Spin-orbit coupling". **c,** Fermi surfaces of two bands (upper: electron; lower: hole) under spin-orbit coupling calculations. Different colors indicate different parts of the Fermi surface in the Brillouin zone. **d,** Temperature dependences of the longitudinal electric resistivity ($\rho$) in zero and 9-T fields. In zero field, a residual resistivity ratio (RRR, $\rho_{300K}/\rho_{2K}$) value of 8.8 and a residual resistivity of $\rho_{2K} \sim 50$ μΩ cm is observed; $\rho_{300K}$ and $\rho_{2K}$ are resistivities at 300 K and 2 K, respectively. **e,** Magnetoresistance measured in fields up to 14 T at 2 K, showing a non-saturated positive magnetoresistance. **f,** Hall data with a non-linear behaviour at high fields, indicating the coexistence of electron and hole carriers at 2 K. All transport measurements depicted here were performed in out−of−plane configuration with *I* // *x* // $[2\bar{1}\bar{1}0]$ and *B* // *z* // $[0001]$. The *x* and *z* axes, respectively, are thus parallel to the *a* and *c* ones shown in **a**. The hexagon in the inset to **d** shows the crystallographic orientations of the crystal. The insets to **d** (right inset), **e** and **f** show the directions of the current and magnetic field in the measurements.

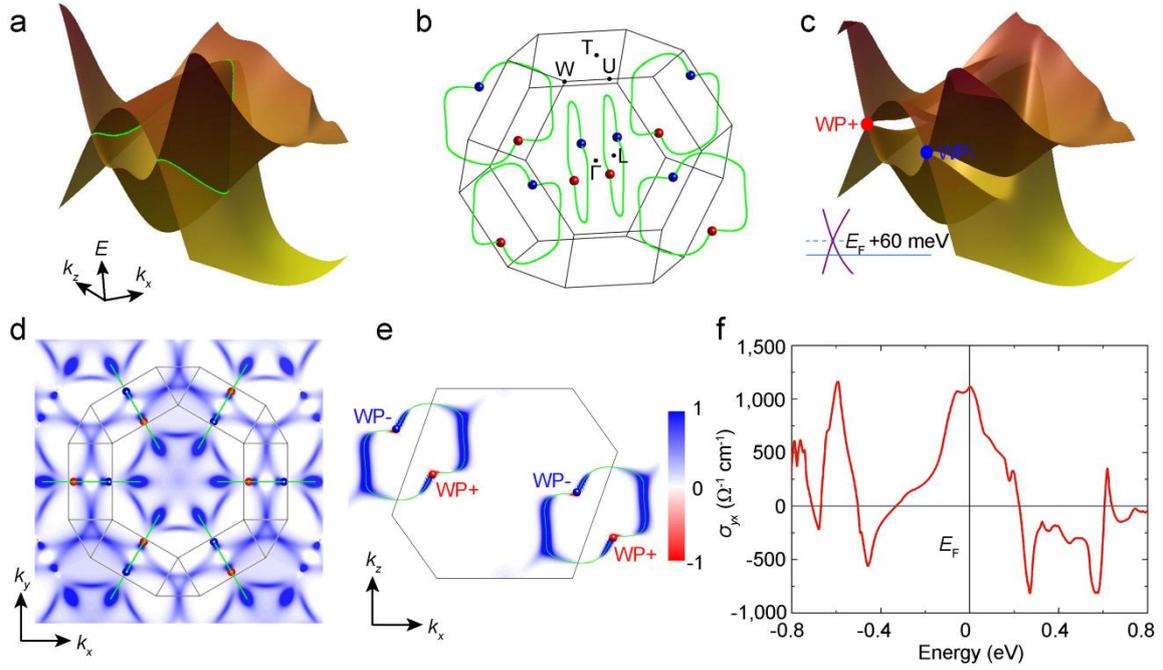

**Figure 2 | Theoretical calculations of the Berry curvature and anomalous Hall conductivity. a**, Linear band crossings form a nodal ring in the mirror plane. **b,** The nodal rings and distribution of the Weyl points in the Brillouin zone. **c**, Spin-orbit coupling breaks the nodal ring band structure into opened gaps and Weyl nodes. The Weyl nodes are located just 60 meV above the Fermi level, and the gapped nodal lines are distributed around the Fermi level. **d,** Berry curvature distribution projected to the $k_x$–$k_y$ plane. **e,** Berry curvature distribution in the $k_y = 0$ plane. The color bar for **d** and **e** are in arbitrary units. **f,** Energy dependence of the anomalous Hall conductivity in terms of the components of $\Omega^z_{yx}(k)$.

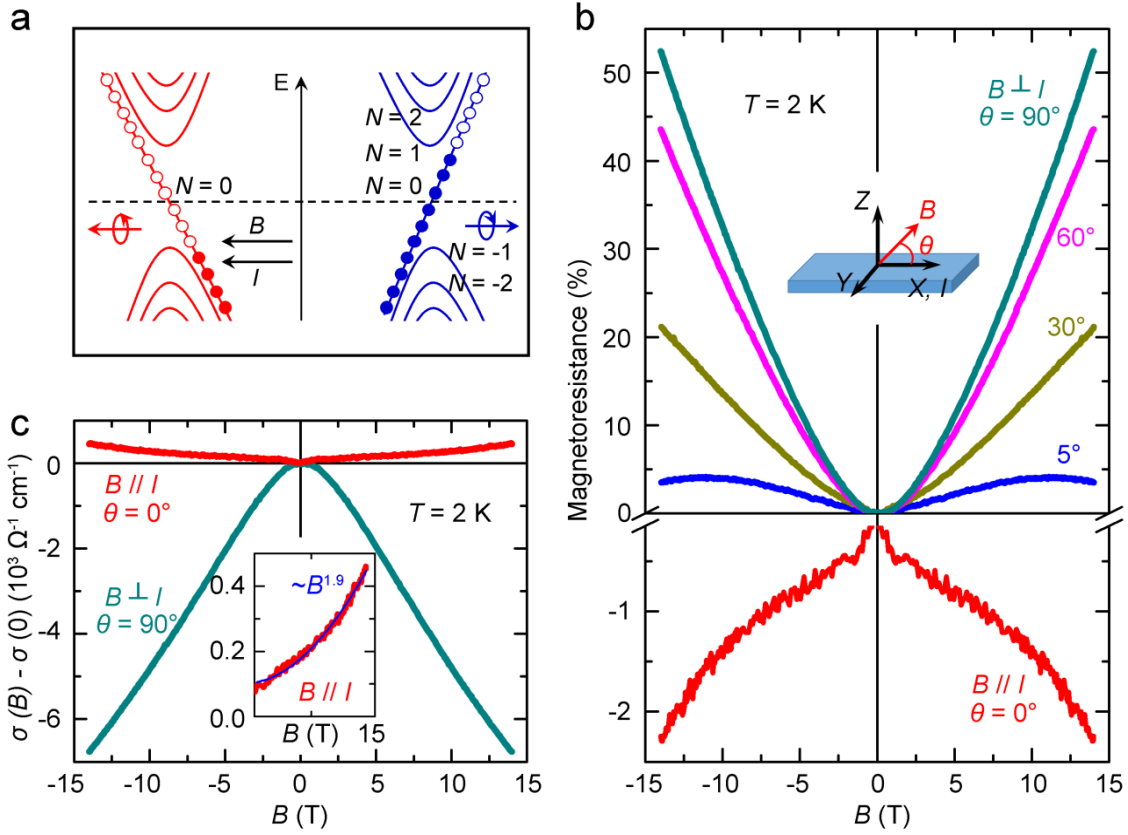

**Figure 3 | Chiral anomaly induced negative magnetoresistance. a**, Schematic of chiral anomaly. When electron current *I* and magnetic field *B* are not perpendicular, the charge carriers pump from one Weyl point to the other one with opposite chirality, which leads to an additional contribution to the conductivity and negative magnetoresistance. **b**, Angle dependence of magnetoresistance at 2 K. For $B \perp I // x // [2\bar{1}\bar{1}0]$ ($\theta$ = 90 °), the magnetoresistance curve shows a positive, non-saturated behaviour up to 14 T. The MR decreases rapidly with decreasing angle. A negative magnetoresistance appears when $B // I // x // [2\bar{1}\bar{1}0]$ ($\theta$ = 0 °). A schematic diagram of the sample geometry is shown for the configuration. **c**, Magnetoconductance at 2 K in both cases of $B \perp I$ and $B // I$. The magnetoconductance is an equivalent description for the magnetoresistance. The positive magnetoconductance is observed in $Co_3Sn_2S_2$ when $B // I$. The fitting of the positive magnetoconductance in Inset shows a ~$B^{1.9}$ dependence, which is very close to the parabolic (~$B^2$) field dependence for Weyl fermions.

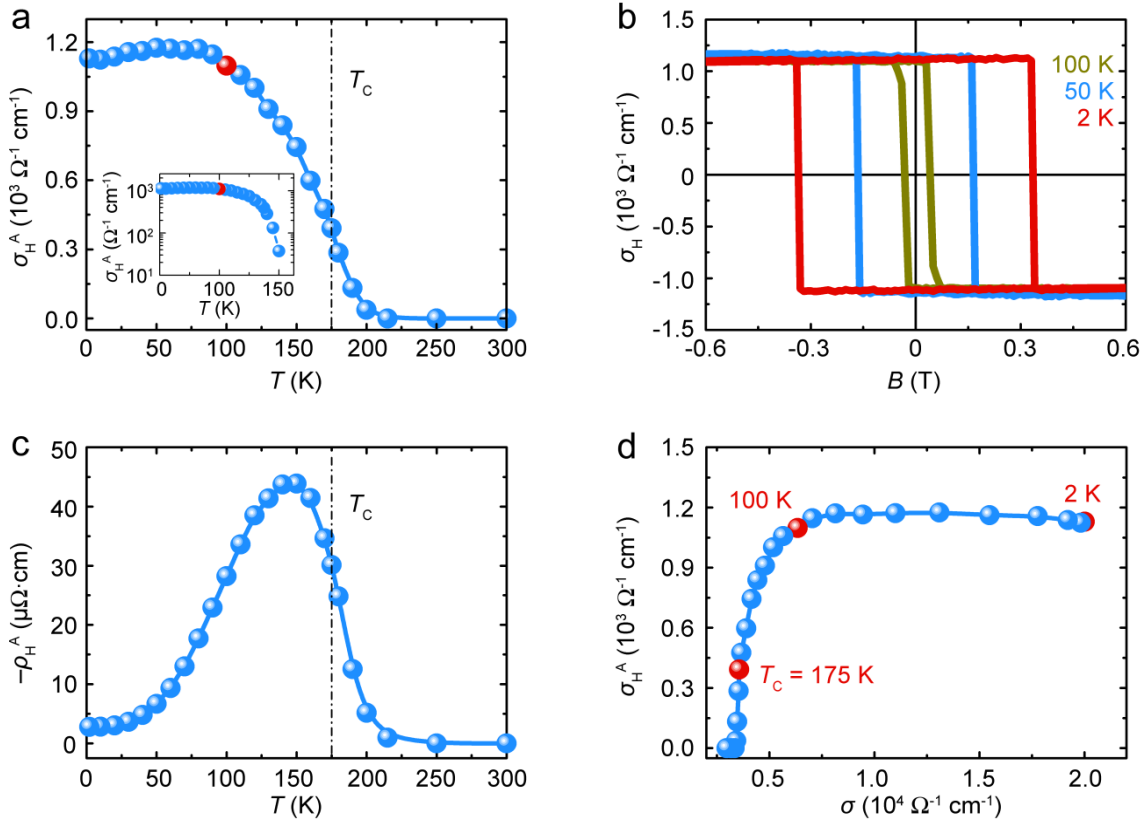

**Figure 4 | Transport measurements of the AHE. a**, Temperature dependence of the anomalous Hall conductivity ($\sigma_H^A$) at zero magnetic field. The inset shows the logarithmic temperature dependence of $\sigma_H^A$. **b**, Field dependence of the Hall conductivity $\sigma_H$ at 100, 50, and 2 K with $I // x // [2\bar{1}\bar{1}0]$ and $B // z // [0001]$. Hysteretic behaviour and the sharp switching appears at temperatures below 100 K. **c**, Temperature dependence of the anomalous Hall resistivity ($\rho_H^A$). A peak appears around 150 K. Since $\rho_H^A$ was derived by extrapolating the high-field part of $\rho_H$ to zero field, non-zero values can be observed just above $T_C$ due to the short-range magnetic exchange interactions enhanced by high fields. **d**, $\sigma$ dependence of $\sigma_H^A$. The $\sigma$-independent $\sigma_H^A$ (i.e., $\sigma_H^A \sim (\sigma)^0$ = constant), below 100 K, puts this system into the intrinsic regime according to the unified model of AHE physics (for more details see Supplementary Information)[39,40].

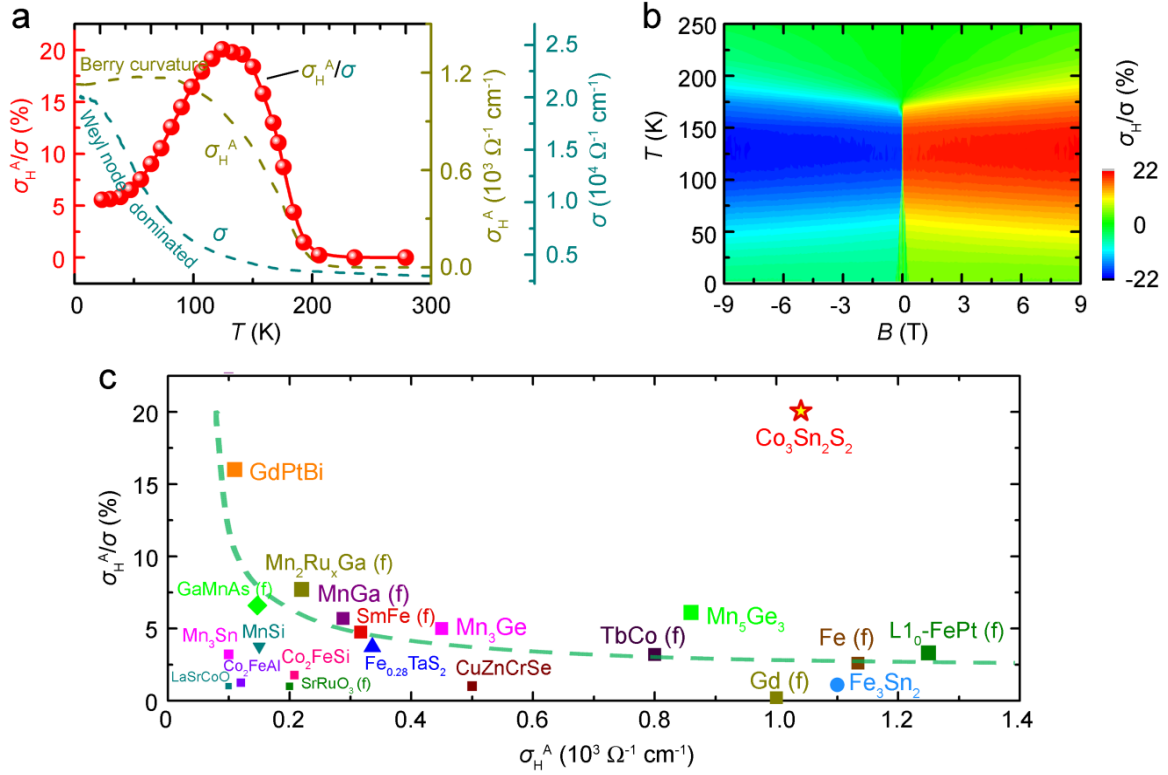

**Figure 5 | Transport measurements of the anomalous Hall angle. a**, Temperature dependences of the anomalous Hall conductivity ($\sigma_H^A$), the charge conductivity ($\sigma$), and the anomalous Hall angle ($\sigma_H^A/\sigma$) at zero magnetic field. Since the ordinary Hall effect vanishes at zero field, only the anomalous Hall contribution prevails (see Supplementary Information). **b**, Contour plots of the Hall angle in the *B–T* space. **c**, Comparison of our $\sigma_H^A$-dependent anomalous Hall angle results and previously reported data for other AHE materials. "(f)" denotes thin-film materials. The dashed line is a guide to the eye. The reported data were taken from references that can be found in the Supplementary Information.